\documentstyle[nato,psfig]{crckapb}

\begin{opening}
\title{Clump stars in the Solar Neighbourhood}

\author{L\'eo Girardi}
\institute{Dipartimento di Astronomia, Universit\`a di Padova\\
           Vicolo dell'Osservatorio 5, I-35122 Padova, Italy}

\end{opening}

\runningtitle{Clump stars in the Solar Neighbourhood}

\begin{document}

\newcommand{\ga}{\stackrel{>}{_{\sim}}}
\newcommand{\la}{\stackrel{<}{_{\sim}}} 
\newcommand{\Msun}{\mbox{$M_\odot$}}
\newcommand{\sub}[1]{\mbox{$_{\rm #1}$}}
\newcommand{\Mhe}{\mbox{$M\sub{cl}$}}
\newcommand{\feh}{\mbox{[Fe/H]}}

\paragraph{Abstract:} 
{\em Hipparcos} data has allowed the identification of a large number 
of clump stars in the Solar Neighbourhood. We discuss our present
knowledge about their distributions of masses, ages, colours,
magnitudes, and metallicities. We point out that the age distribution
of clump stars is ``biased'' towards intermediate-ages. Therefore, the
metallicity information they contain is different from that provided
by the local G dwarfs.  Since accurate abundance determinations are
about to become available, these may provide useful constraints to
chemical evolution models of the local disc.

\subsubsection*{\bf 1. The clump after {\em Hipparcos}}
\label{sec_intro}

The clump of core-helium burning stars has been known for decades in
the colour-magnitude diagrams (CMD) of open clusters and nearby
galaxies. However, only recently we have been able to identify
a significant number of clump stars in the Solar Neighbourhood, thanks
to the {\em Hipparcos} mission (Perryman et al.\ 1997). In fact, the
ESA (1997) catalog contains $\sim600$ clump stars with parallax error
lower than 10~\%, and hence an error in absolute magnitude lower than
0.12~mag. This accuracy limit corresponds to a distance of
$\sim125$~pc within which the sample of clump stars is complete.
Moreover, accurate $BV$ photometry (and $I$ for $\sim1/3$ of the
sample), is available, and interstellar absorption is small enough to
be neglected.

\begin{figure}
\psfig{file=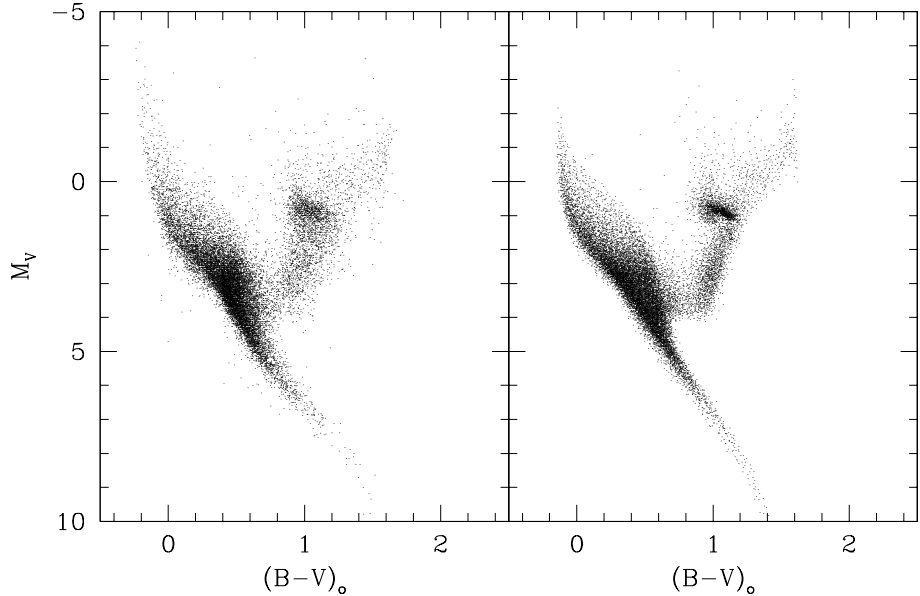,width=\textwidth}
       	\caption{  
Left panel: {\em Hipparcos} data. Right panel: theoretical simulation 
(see text).
       }	 
\label{fig_hipp} 
\end{figure}

The {\em Hipparcos} $M_V$ versus $B-V$ CMD is illustrated on the left
panel of Fig.~\ref{fig_hipp}. In this plot, we exclude binaries and
limit the sample to stars with $\pi<0.007$~arcsec and $V<8.5$. The
clump is well evident at $M_V\simeq1$, $B-V\simeq 1$.

So far, most of the emphasis on nearby clump stars has been on their
possible use as standard candles, since their mean absolute magnitude
can be determined with an accuracy of hundredths of magnitude from
{\em Hipparcos} database (cf.\ Paczy\'nski \& Stanek 1998; Stanek et
al.\ 1998). This application requires a good understanding of the
basic characteristics of clump stars, in order to access the
systematic changes of the clump luminosity in different populations,
as those pointed out by Cole (1998) and Girardi et al.\ (1998).

\subsubsection*{\bf 2. The mass distribution}
\label{sec_mass}

\begin{figure}
\centering{\psfig{file=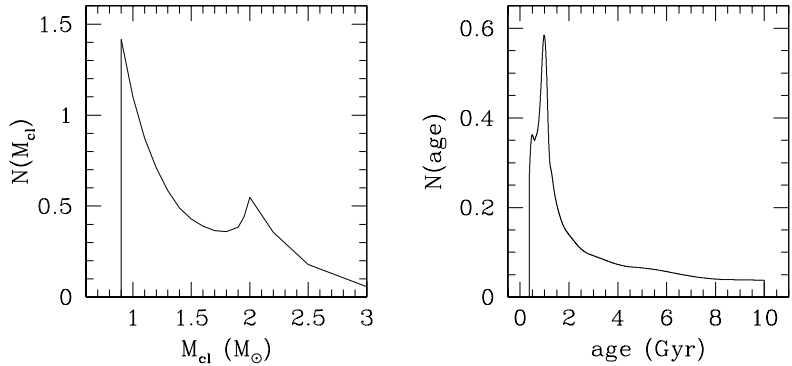,width=0.85\textwidth}}
       	\caption{
	Mass (left panel) and age (right panel) distribution of 
clump stars for the case of a constant star formation rate (see text). 
       }	 
\label{fig_mass} 
\label{fig_age} 
\end{figure}

A common idea is that the mass distribution of clump stars roughly 
follows the IMF, presenting then a single and sharp peak at the 
lowest possible masses, i.e.\ with $0.8-1.2$~\Msun. This reasoning, 
however, is not valid for evolved stars such as core-helium
burners. From simple arguments we can derive their mass distribution,
in a galaxy of age $T$, as being proportional to the IMF $\phi_M$, to
the core-helium burning lifetime $t\sub{He}$, and to the star
formation rate (SFR) at the epoch of birth $\psi[T-t(M)]$.  Since
$t\sub{He}$ presents a peak at about 2~\Msun\ (the transition from
low- to intermediate masses), and the IMF shows a peak at the lowest
masses, a {\em double-peaked mass distribution} appears (Girardi
1999). This is shown at the left panel of Fig.~\ref{fig_mass}, for the
case of a constant SFR from 0.1 to 10~Gyr ago, and a Salpeter IMF.

Moreover, it turns out that some intermediate-mass stars, from say 2
to 2.5~\Msun, are not severely under-represented with respect to the
low-mass ones. They are close enough to the clump region in the CMD to
be considered as genuine clump stars. For the case of a constant SFR,
they make about 20~\% of the clump.

However, is the hypothesis of a constant SFR a realistic one for the 
local disc\,? From careful 
analyses of the {\em Hipparcos} CMD, Bertelli \& Nasi (1999) favour an 
almost-constant SFR with {\em increasing} rates in the last few 
Gyr, whereas Gilmore et al.\ (this meeting) find several
episodes of star formation, without any marked long-term trend 
of decreasing or increasing SFR. An independent indication for a 
roughly-constant SFR (over time-scales of several Gyr) comes from the 
simulation we present in the right
panel of Fig.~\ref{fig_hipp}. Using a population synthesis code, we
generate an uniform distribution of stars, and limit it to
$\pi<0.007$~arcsec and $V<8.5$.  The stars are picked up from the
Girardi et al.\ (2000) set of evolutionary tracks and isochrones, and
distributed according to a constant SFR and with a metallicity
distribution $\feh=-0.12\pm0.18$ (see \S~6 below). 
Although this was intended to be a {\em crude} initial model
for the local stars, the derived synthetic CMD reproduces quite
well the observed number of stars in several key regions of the {\em
Hipparcos} CMD (Girardi et al., in preparation). 

\subsubsection*{\bf 3. The age distribution}
\label{sec_age}

A constant SFR does not mean a constant age distribution, at least not
for evolved stars. In fact, from the mass distribution of clump stars
obtained with a constant SFR, we can easily
derive the age distribution shown in the right panel of Fig.~\ref{fig_age}.
The latter is far from being constant; it peaks at an age of $1$~Gyr, and
decreases monotonically afterwards. Half of the clump stars have ages
lower than 2~Gyr\,! This result comes, essentially, from the
continuous decrease, with age, of the birth rate of post-main sequence
stars.

Therefore, the age distribution of clump stars (K giants) is very 
different from that of low-main sequence stars (G dwarfs).  The 
long-lived G dwarfs have an age distribution simply proportional to the 
SFR, whereas K giants have it ``biased'' towards
intermediate-ages ($1-3$~Gyr).  This difference reflects also into the
metallicities: since younger stars are more metal rich, giants tend to
have a relatively narrow metallicity distribution, if compared to
dwarfs (see \S~6 below).

These results are in contrast with a common prejudice, i.e.\ that red
giants sample equally well stellar populations of all ages, from the
intermedia\-te-age (metal rich) to the oldest (metal-poor) ones in a
galaxy.  This is not the case. Only in galaxies which formed few stars
at intermediate-ages, can the age distribution of red giants be
expected to be moderately flat.

\subsubsection*{\bf 4. The colour distribution}
\label{sec_colour}

Red giant stars are known to become redder at higher metallicities.
Jimenez et al.\ (1998) applied this concept to derive, solely from the
colour range of {\em Hipparcos} clump stars, an estimate of
metallicity range covered by them, obtaining $-0.7<\feh<0.0$. Although
this result has little effect on their estimate of the age of the 
Galactic disc (Carraro, this meeting), the approach is essentially 
misleading. In
fact, red giants become redder also at lower masses, or equivalently,
at higher ages.  Girardi et al.\ (1998) demonstrated that a galaxy model
with mean solar metallicity, a {\em very small} metallicity dispersion
($\sigma_{\rm [Fe/H]}=0.1$~dex), and constant SFR up to 10~Gyr ago, 
shows a clump as wide in colour as the observed {\em Hipparcos} one -- 
i.e.\ with $\Delta(V-I)\simeq0.2$~mag (see also Fig.~\ref{fig_hipp}).
It follows that a significant fraction of the colour spread of the 
local clump might be due to an age spread, rather than to a metallicity
spread.

\subsubsection*{\bf 5. The magnitude distribution}
\label{sec_mag}

The correct interpretation of the colour spread of clump stars is also
crucial for distance determinations. Were the clump colour determined
by metallicity only, the observed constancy of the $I$-band magnitude
with colour inside the clump, in different galaxies, would be
indicating that $M_I$ is virtually independent of metallicity, and
hence an excelent standard candle (cf.\ Paczy\'nski \& Stanek 1998;
Stanek et al.\ 1998; Udalski 1998a).

This conclusion is not supported by theoretical models (Cole 1998;
Girardi et al.\ 1998), which predict a brighter clump at lower
metallicities, and a more complex dependence on age. The observational
works by Udalski (1998a,b), instead, conclude for a very modest
dependence of $M_I$ on both metallicity and age.  However, the data
analyses contain several uncertain assumptions, like the large
``geometric corrections'' applied to the distances of LMC and SMC
clusters in Udalski's (1998b) study.  More recently, Twarog et al.\
(1999) and Sarajedini (1999) concluded for a larger age and
metallicity dependence of the clump magnitude, using data from disc
open clusters.

Models predict that the clump magnitude depends on age in a
non-monotonic way, being fainter than the mean (by up to 0.4~mag) at
both the $\sim1$~Gyr and $\ga10$~Gyr age intervals. These two groups 
with the lowest luminosities correspond also to the bluest and reddest 
clump stars one finds for a given metallicity. The effect is such that,
summing up stars of all ages, one can easily obtain a
nearly-horizontal clump in the $M_I$ versus $V-I$ CMD, similar to the
{\em Hipparcos} one (Girardi et al.\ 1998). Almost-horizontal
structures can also be obtained in galaxy models with decreasing SFRs
and a reasonable age-metallicity relation (cf.\ Girardi 1999). Thus,
the observed constancy of $M_I$ with colour in the clump of different
galaxies can be reasonably reproduced by models.  Nonetheles, there is 
no reason for $M_I$
being constant from galaxy to galaxy. In fact, the LMC clump should be
intrinsically brighter than the local one by $0.2-0.3$~mag in the
$I$-band (Cole 1998; Girardi et al.\ 1998).

A fine structure in the CMD -- namely a fainter secondary clump plus a
low-density bright plume -- comes out in galaxy models containing
$\sim1$-Gyr old populations with metallicities $Z\ga0.004$ (Girardi et
al.\ 1998; Girardi 1999). These features are due to the
intermediate-mass clump stars, i.e.\ those just massive enough for
starting to burn helium in non-degenerate conditions. A nice
confirmation of the predictions is provided by Bica et al.\ (1998) and
Piatti et al.\ (1999), who detected a group of faint clump stars in
several fields over the LMC. Similar structures are also present in
the {\em Hipparcos} CMD (Girardi et al.\ 1998; Beaulieau \& Sackett
1998).

\subsubsection*{\bf 6. The metallicity distribution} 
\label{sec_metal}

The metallicities of 581 nearby K giants (mainly clump
stars) have been derived by H\o g \& Flynn (1998), based on DDO
photometry.  From their data it turns out that the $V-I$ colour does
not correlate with \feh\ (cf.\ Paczy\'nski 1998), contrarily to what
expected, for intance, from Jimenez et al.\ (1998) analysis of local
clump stars.  On the basis of the arguments discussed in \S~4 above, 
however, this result is not surprising: it might simply reflect that 
the colour spread of the local clump is mostly due to an age spread 
(from $\sim1$ to 10 Gyr), rather than due to a metallicity spread.

Girardi \& Salaris (in preparation) selected a sample of clump stars
with spectroscopic abundance determinations.  The resulting
metallicity distribution is Gaussian-like with a very small
dispersion, i.e.\ $\feh=-0.12\pm0.18$~dex.  This disagrees with the
metallicity range ($-0.7\le\feh\le0.0$) derived from the colours by
Jimenez et al.\ (1998).  Again, the spectroscopic \feh\ determinations
indicate that the colour spread in the local clump cannot be simply
due to a spread of metallicity.

Another evident aspect in the metallicity distribution of clump stars
is the anallogous of the G-dwarf problem: only 15 out of 334 clump
stars have $\feh<-0.5$, whereas {\em at least} 43 are expected from a
simple closed-box model of chemical evolution. This new ``K-giant
problem'' is quantitatively different from the G-dwarf one, since it
is offered by stars which are, on the mean, younger than G
dwarfs. Therefore, the chemical information provided by clump stars
should be considered as additional to that provided by the dwarfs, and
more suitable to probe the chemical conditions at intermediate-ages
(from say 1 to 4 Gyr ago).

\subsubsection*{\bf 7. Concluding remarks}

The {\em Hipparcos} catalog provides us an extremely interesting 
sample of clump stars, complete up to 125~pc, and representing a 
well-understood evolutionary stage. Thus far, it has provided
interesting checks to the theory of stellar evolution and population
synthesis. A good example is the finding of a fine structure in the 
clump, which represents a sort of ``lower main sequence'' for core-helium 
burning stars. Theoretical modelling of the clump in galaxies also 
brings some unexpected results, such as the wide and double-peaked 
mass distribution, and the intrinsic bias towards intermediate-ages. 
These aspects should be taken into account when interpreting the 
observational data related to clump stars, and to red giants in 
general.

The perspectives of using the clump data for probing the chemical
evolution of the Galactic disc are also very promising. Grenon and
Morossi (both in this meeting) announced more reliable and 
homogeneous chemical abundance determinations for nearby red giants.
With these new data, the sample of nearby clump stars may render 
additional (and quantitative) constraints to chemical evolution 
models of the Solar Neighbourhood. In this respect, we should also 
consider the tight constraints to the disc SFR provided by the
detailed analyses of the {\em Hipparcos} CMD.


\begin{thebibliography}{} 
\bibitem{} Beaulieu J.-P., Sackett P.D., 1998, AJ 116, 209
\bibitem{} Bertelli G., Nasi E., 1999, A\&A in press
\bibitem{} Bica E., Geisler D., Dottori H., et al., 1998, AJ 116, 723
\bibitem{} Cole A.A., 1998, ApJ 500, L137
\bibitem{} ESA, 1997, The Hipparcos and Tycho Catalogues, ESA SP-1200
\bibitem{} Girardi L., Groenewegen M.A.T., Weiss A., Salaris M., 1998,
	 MNRAS 301, 149
\bibitem{} Girardi L., 1999, MNRAS 308, 818
\bibitem{} Girardi L., Bressan A., Bertelli G., Chiosi C., 2000, A\&AS
	in press
\bibitem{} H\o g E., Flynn C., 1998, MNRAS 294, 28
\bibitem{} Jimenez R., Flynn C., Kotoneva E., 1998, MNRAS 299, 515
\bibitem{} Paczy\'nski B., 1998, Acta Ast.\ 48, 405 
\bibitem{} Paczy\'nski B., Stanek K.Z., 1998, ApJ 494, L219 
\bibitem{} Perryman M.A.C., Lindegren L., Kovalevsky J., et al.,
	1997, A\&A 323, L49
\bibitem{} Piatti A., Geisler D., Bica E., et al., 1999, AJ in press
	(astro-ph/9909475)
\bibitem{} Sarajedini A., 1999, AJ in press
\bibitem{} Stanek K.Z., Zaritsky D., Harris J., 1998, ApJ 500, L141
\bibitem{} Udalski A., 1998a, Acta Astr.\ 48, 113
\bibitem{} Udalski A., 1998b, Acta Astr.\ 48, 383
\end{thebibliography}
\end{document}